\documentclass[12pt,eqsecnum]{revtex4}
\begin{document}
\title{Various forms of BRST symmetry in Abelian 2-form gauge theory}


\author{ Sumit Kumar Rai \footnote{e-mail address: sumitssc@gmail.com}}
\author{Bhabani Prasad Mandal \footnote{e-mail address:
 \ \ bhabani.mandal@gmail.com, \ \  bhabani@bhu.ac.in  }}


\affiliation{ Department of Physics,\\
Banaras Hindu University,\\
Varanasi-221005, INDIA. \\
}

\begin{abstract}
We derive the various forms of BRST symmetry using Batalin-Fradkin-Vilkovisky approach in the 
case of
Abelian 2-form gauge theory. We show that the so-called dual BRST 
symmetry is not an independent symmetry but the generalization of BRST symmetry obtained from the canonical transformation 
in the bosonic and ghost sector. We further obtain the new forms of both BRST and dual-BRST symmetry 
by making a general transformation in the Lagrange multipliers of the bosonic and ghost sector of the theory.
\end{abstract}
\maketitle
\section{Introduction}
Abelian antisymmetric gauge theories have become the subject of interest in various aspects. 
It was first shown by Kalb and Ramond that Abelian rank-2 antisymmetric fields can interact 
with classical strings \cite{kara}, which was further applied to the dual formulation of 
Abelian Higgs model \cite{suga}. Antisymmetric tensor field appear coupled to gravity or 
supergravity fields with higher curvature term in four and ten dimensions \cite{sase} and 
complete understanding of these couplings in superstring theories are crucial in order to 
have anomalies cancellation \cite{grsc}. The rank-2 antisymmetric tensor field generates 
effective mass for an Abelian vector gauge field through a topological coupling between these 
two fields \cite{crsc}. A geometric aspect of Abelian rank-2 antisymmetric tensor fields has 
been discussed in a U(1) gauge theory loop space \cite{frne}. A superspace formulation of Abelian antisymmetric gauge theories and the time evolution invariance of the celebrated Curci-Ferrari type of restrictions \cite{cufe} invoked in Abelian 2-form gauge theory to obtain an absolutely anticommuting and off-shell nilpotent (anti-) BRST as well as (anti-) co-BRST symmetry transformation have also been studied \cite{dema,mama}.
BRST formalism \cite{berost} is one of the most elegant method of covariant canonical quantization of gauge theories. Using BRST formalism, the canonical covariant quantization of an Abelian rank-2 antisymmetric tensor field has been studied  by many authors \cite{kaul,frto} in a systematic manner.

A general method of Hamiltonian BRST quantization of systems with first class constraints was developed by Batalin-Fradkin-Vilkovisky (BFV)\cite{frvi}. BFV approach does not require closure off-shell of the gauge algebra and therefore does not need an auxiliary field. This formalism heavily relies on BRST transformations which are independent of the gauge condition. Being based on Hamiltonian, the approach is closer to Hilbert space techniques and unitarity. This method uses an extended phase space where the Lagrange multipliers and the ghosts are treated as dynamical variables. The generator of the BRST symmetry for systems with first class constraints can be constructed from the constraints in a gauge independent way whose cohomology produces the physical states. The BFV technique has been widely studied on various model with first class constraints \cite{gara,nega,gaete,rama} and  second class constraints \cite {kipa,kiki}. 

In 2-form gauge theories, the framework of BFV approach has been used to show the quantum 
equivalence of massless antisymmetric second rank tensor field theories with massless scalar 
and gauge theories in four and five dimensions \cite{risa}, to obtain the covariant 
representations for the S-matrix of a self-interacting antisymmetric tensor field \cite 
{rybk}. It has also been applied to obtain the operator of BRST transformations and the BRST 
invariant effective Lagrangian of an "Interacting antisymmetric tensor field" using 
Hamiltonian BFV-BRST quantization of systems with dependent first class constraints \cite
{frlo}.

BRST symmetry provides a basis for the modern quantization of gauge theories. Various forms 
of BRST symmetry have been studied in 1-form gauge theory such as non-local and non-covariant 
BRST \cite{lama}, covariant and non-local BRST \cite{tafi}, non-covariant and local  BRST 
symmetry \cite{yale} and another local, covariant and off-shell nilpotent BRST symmetry \cite
{rive,lahi}.
  It has also been shown 
that there exists a local, covariant and nilpotent BRST symmetry, the so-called dual -BRST 
symmetry under which the gauge fixing term remains invariant for a free U(1) gauge theory and 
QED \cite{malik1,malik2}.
In 2-form gauge theory, another  local, covariant and nilpotent dual-BRST symmetry  is claimed to be the independent symmetry and not the 
generalization of BRST symmetry, where the generator of this symmetry is analogous to the 
co-exterior derivative of the differential geometry\cite{hama,guma}.

In this paper, we investigate whether the  dual-BRST symmetry is a new independent 
symmetry or it is merely an artifact of the canonical transformation in the bosonic sector 
and ghost sector. We also derive new forms of BRST and dual-BRST symmetry by making a general 
transformation of the Lagrange multipliers in both bosonic and ghost sectors.

The outline of this paper is as follows. In Sec. II, we derive the BRST symmetry using BFV approach for the case of Abelian 2-form gauge theory. In Sec. III, we discuss dual-BRST symmetry obtained from the canonical transformations  given in Sec. III-A. In Sec. IV and V, we obtain a new form of BRST symmetry and dual-BRST symmetry respectively. Finally, in Sec. VI, we make  concluding remarks and enlight some new directions for future works.

\section{BFV-BRST Approach}
We start with the Abelian free 4-D Kalb-Ramond Lagrangian density \cite{kara} as
\begin{equation}
{\cal{L}}_0=\frac{1}{12} F^{\mu\nu\lambda} F_{\mu\nu\lambda},
\end{equation}
where the antisymmetric field tensor is defined as $F_{\mu\nu\lambda}=\partial_\mu 
B_{\nu\lambda}+ \partial_\nu B_{\lambda\mu}+ \partial_\lambda B_{\mu\nu}$. In order to find 
the Hamiltonian, we calculate the canonical momenta conjugate to $B_{0i}$ and $B_{ij}$ as
\begin{eqnarray}
\Pi^{0i}&=&\frac{\partial{\cal{L}}_0}{\partial\dot{B}_{0i}}=0,\\
\Pi^{ij}&=&\frac{\partial{\cal{L}}_0}{\partial\dot{B}_{ij}}=\frac{1}{2} F^{0ij}.
\end{eqnarray}

Using Dirac's prescriptions \cite{pam} for constraint analysis, we note that the primary 
constraint in the theory is
\begin{equation}
\Pi^{0i}\approx 0,\quad \quad i=1,2,3. \label{cons1}
\end{equation}
and
\begin{equation}
\partial_i\Pi^{ij}\approx 0,\label{cons2}
\end{equation}
is the secondary constraint. It can be checked that there are no further constraints and the 
constraint given in Eq. (\ref {cons1}) and Eq. (\ref{cons2}) are first class constraints \cite
{kaul}. The canonical pairs $(B_{0i},\Pi^{0i})$ and $(B_{ij},\Pi^{ij})$ in the original phase 
space satisfy the following canonical commutation relations
\begin{eqnarray}
\left[B_{0i}({\bf x}),\; \Pi^{0j}({\bf y})\right] &=&\;i \;\delta_i^j \;\delta^3 ({\bf x}-{\bf
 y}),\label{com1}\\
\left[B_{ij}({\bf x}),\; \Pi^{kl}({\bf y})\right] &=&\;\frac{i}{2}\;(\delta_i^k\delta_j^l -
\delta_j^k\delta_i^l)\; \delta^3 ({\bf x}-{\bf y}).\label{com2}
\end{eqnarray}
The canonical Hamiltonian  is given by
\begin{eqnarray}
H_c&=&\int d^3x \left(\Pi^{\alpha\beta}\dot{B}_{\alpha\beta}-{\cal{L}}_{kin}\right) \nonumber\\
&=&\int d^3x\;\left(\Pi_{ij}\Pi_{ij}+\frac{1}{12}F_{ijk}F_{ijk}+2\Pi^{ij}\partial_i B_{0j}
\right)\nonumber\\
&=&\int d^3x\;\left({\cal{H}}_0+\lambda\Phi \right),
\label{h0}
\end{eqnarray}
where, $\lambda \sim B_{0i}$ (in this case) is the Lagrange multiplier associated with the 
secondary constraint $\Phi =\partial_i\Pi_{ij}$ and $H_0=\int d^3x{\cal{H}}_0=\int d^3x\; 
\left(\Pi_{ij}\Pi_{ij}+\frac{1}{12}F_{ijk}F_{ijk}\right)$.
Using BFV approach \cite{frvi}, we extend the original phase space by 
introducing the pair of canonically conjugate anticommuting ghosts $({\cal{C}}_i,{\cal{P}}_i)$
, $(\bar{{\cal{C}}}_i,\bar{{\cal{P}}}_i)$ and the pair of canonically conjugate commuting 
ghosts $(\beta, \Pi_\beta)$, $({\bar{\beta}},\Pi_{\bar{\beta}})$ for each of the first class 
constraints. They have the following ghosts numbers
\begin{eqnarray}
gh \;{\cal{C}}_i&=&-gh\; {\cal{P}}_i=1, \quad \quad gh \;\bar{{\cal{C}}}_i=-gh \;\bar{{\cal
{P}}}_i=-1,
\label{ghn}\\
gh \;\beta &=&-gh \; \Pi_\beta =2, \quad \quad gh\;\bar{\beta}=-gh \;\Pi_{\bar{\beta}}=-2.
\end{eqnarray}
They satisfy the following (anti-)commutation relations
\begin{eqnarray}
\left\{{\cal{C}}_i({\bf x}),{\cal{P}}_j({\bf y}) \right\}&=&\; -i\; \delta_{ij}\;\delta^3 ({
\bf x}-{\bf y}), \quad \quad \left\{\bar{{\cal{C}}}_i({\bf x}),{\bar{\cal{P}}}_j({\bf y}) 
\right\}=
\; -i\; \delta_{ij}\delta^3 ({\bf x}-{\bf y}),  \label{acom1}\\
\left[ \beta({\bf x}),\; \Pi_\beta ({\bf y})\right ]&=&\; i\delta^3 (\bf{x}-\bf{y}),\quad 
\quad \left[ \bar{\beta}({\bf x}),\;\Pi_{\bar{\beta}} ({\bf y})\right] =\; i\;\delta^3 ({\bf 
x}-{\bf y}).\label{com3}
\end{eqnarray}
The phase space is further extended by introducing canonical conjugate pairs 
$({\cal{C}}_0,{\cal{P}}_0)$ and (${\bar{\cal{C}}}_0,\bar{\cal{P}}_0)$ as Lagrange multipliers 
to the pair $({\cal{C}}_i,{\cal{P}}_i)$, $(\bar{{\cal{C}}}_i,\bar{{\cal{P}}}_i)$ and a 
canonical pair ($\varphi_1,\Pi_{\varphi_1})$ as Lagrange multiplier to the gauge condition. 
They 
satisfy the similar anticommutation relation given in Eq.(\ref{acom1}) and commutation 
relation given in Eq. (\ref{com3}) respectively. They have the ghost number as mentioned in 
Eq. (\ref{ghn}).

The effective action in the extended phase space becomes
\begin{eqnarray}
S_{eff}&=&\int d^4x \left[ \Pi^{0i}\dot{B_{0i}}+\Pi^{ij}\dot{B_{ij}}+\dot{\cal{C}}_i{\cal
{P}}_i+\dot{\bar{\cal{C}}}_i\bar{\cal{P}}_i+\dot{\cal{C}}_0{\cal{P}}_0+\dot{\bar{\cal{C}}}_0
\bar{\cal{P}}_0+\Pi_\beta \dot\beta \right . \nonumber\\
&+&\left .\Pi_{\bar{\beta}}\dot{\bar{\beta}}+\Pi_\varphi\dot\varphi -{\cal{H}}_
\Psi\right],\label{seff}
\end{eqnarray}
where 
\begin{equation}
H_\Psi=\int d^3x{\cal{H}}_\Psi= H_0+\left\{Q,\Psi\right\}.
\end{equation}
$\Psi $ is the gauge fixed fermion and Q is the generator of the BRST symmetry. The form of 
the symmetry generator in the extended phase space is
\begin{equation}
Q_b=i\int d^3x\left[2\partial_i\Pi_{ij}{\cal{C}}_j+\Pi_\varphi{\bar{\cal{P}}}_0-{\cal{P}}_0
\Pi_{\bar
{\beta}}+\bar{\cal{P}}_i\Pi_{0i}+{\cal{P}}_i\partial_i\beta \right ],\label{brsc}
\end{equation} 
which satisfies the following algebras
\begin{eqnarray}
\left\{Q_b,Q_b\right\}&=&0, \quad \quad \left\{H_\Psi,Q_b\right\}=0,\nonumber\\
gh(Q_b)&=&1,\quad \quad gh(H_\Psi)=0.
\end{eqnarray}
Using the relation $\delta_b\phi=-i{\left[\phi,Q_b\right]}_\pm $ (+ sign for fermionic and - for 
bosonic nature of $\phi$), the BRST charge given in Eq. (\ref{brsc}) will generate the 
following BRST transformations
\begin{eqnarray}
&&\delta_b B_{0i}={\bar{\cal{P}}}_i, \quad \quad \quad \delta_b B_{ij}= \left(\partial_i {\cal
{C}}_j-\partial_j{\cal{C}}_i\right), \quad\quad \delta_b \Pi_{\varphi_1}=0, \nonumber\\
&&\delta_b{\cal{C}}_i\;\;=\partial_i \beta, \quad\quad \;\;\delta_b{\bar{\cal{C}}}_i\;\;=\Pi_
{0i}, \quad \quad\quad \quad\quad\;\; \;\;\;
 \delta_b{\cal{C}}_0\;\;=\Pi_{\bar{\beta}},\nonumber\\
&&\delta_b{\bar{\cal{C}}}_0\;=\Pi_\varphi,\quad\quad \quad\delta_b\varphi_1\;=-{\bar{\cal{P}}}_0,
\quad\quad \quad\quad\quad\;\;\;
\delta_b\beta\quad=0,\nonumber\\
&&\delta_b\bar{\beta}\;\;=-{\cal{P}}_0,\quad\;\;\;\;\;\delta_b\Pi_{0i}=0,\quad\quad\quad\quad\; 
\quad\quad\quad \delta_b\Pi_{ij}\;=0,\nonumber\\
&&\delta_b{\cal{P}}_i  =2\partial_j\Pi_{ji},\quad\;\;\; \delta_b{\bar{\cal{P}}}_i\;=0,
\quad\quad\quad\quad\quad\quad\quad\;\delta_b{
\cal{P}}_0\;\;=0,\nonumber\\
&&\delta_b{\bar{\cal{P}}}_0=0,\quad\quad \quad\;\;\;\delta_b\Pi_\beta\;=-\partial_i{\cal{P}}_i,
\quad\quad \quad\quad\quad
\delta_b\Pi_{\bar{\beta}}\;\;=0.\label{qbrst}
\end{eqnarray}
We choose gauge fixed fermion as 
\begin{eqnarray}
\Psi &=&\int d^3y\left[-{\bar{\cal{C}}}_j\partial_iB_{ij}+\frac{1}{2}{\bar{\cal{C}}}_i\Pi_
{0i}+\frac{1}{2}{\bar{\cal{C}}}_0
\Pi_{\varphi_1}-{\cal{P}}_i B_{0i}-{\bar{\cal{C}}}_0\partial_iB_{0i}\right. \nonumber\\
&+&\left. \Pi_\beta{\cal
{C}}_0-\bar{\beta}\partial_i{\cal{C}}_i+{\bar{\cal{C}}}_i\partial_i\varphi +\bar{\beta}{\bar{
\cal{P}}}_0\right].
\end{eqnarray}
We calculate
\begin{eqnarray}
\left\{Q,\Psi\right\}&=&\int d^3 x\left[\partial_i{\bar{\cal{C}}}_j
\left(\partial_i{\cal{C}}_j-\partial_j{\cal{C}}_i \right)- 2\partial_i\Pi^{ij}B_{0j}
-\Pi_{0i}\partial_j B_{ji}+{\bar{\cal{P}}}_i{\cal{P}}_i+\partial_i{\cal{P}}_i {\cal{C}}_0
\right.\nonumber\\
&+&\left.\partial_i\bar{\beta}\partial_i \beta-{\cal{P}}_0\partial_i{\cal{C}}_i
 - \Pi_{\varphi_1}\partial_i B_{0i}-{\bar{\cal{P}}}_i\partial_i
{\bar{\cal{C}}}_0-{\bar{\cal{P}}}_0\partial_i{\bar{\cal{C}}}_i+\Pi_{0i}
\partial_i{\varphi_1}+\frac{1}{2}\Pi_{0i}\Pi_{0i} \right.\nonumber\\
&+&\left.{\cal{P}}_0{\bar{\cal{P}}}_0 +\frac{1}{2}\Pi_{\varphi_1}^2+\Pi_{\bar{\beta}}
\Pi_\beta\right].\label{qpsi}
\end{eqnarray}
Substituting  above equation in Eq. (\ref{seff}), the generating functional 
can be expressed as
\begin{eqnarray}
Z_\Psi &=&\int {\cal{D}}\chi  \;e^{iS_{eff}}\nonumber\\
&=&\int d\Pi_{0i}\; d\Pi_{ij}\; dB_{0i}\;dB_{ij}\;d{\cal{C}}_i\;d{\bar{\cal{C}}}_i\;d{\cal
{P}}_i\;d{\bar{\cal{P}}}_i\;d{\cal{C}}_0\;d{\bar{\cal{C}}}_0\;d{\cal{P}}_0\;d{\bar{\cal{P}}}_0
\;d\beta\;d\bar{\beta}\;d\Pi_{\beta}\;d\Pi_{\bar{\beta}}\nonumber\\
&&d\varphi \; d\Pi_\varphi \exp \left[i\int d^4x\left\{\Pi^{0i}{\dot{B}}_{0i}+\Pi^{ij}{\dot
{B}}_{ij}+\dot{\cal{C}}_i{\cal{P}}_i+{\dot{\bar{\cal{C}}}}_i{\bar{\cal{P}}}_i+{\dot{\cal{C}}}_0
{\cal{P}}_0+{\bar{\cal{C}}}_0{\bar{\cal{P}}}_0+\Pi_\beta\dot{\beta}\right.\right.\nonumber\\
&+&\left.\left.\Pi_{\bar{\beta}}\dot{\bar{\beta}}+\Pi_\varphi\dot{\varphi}-\Pi_{ij}\Pi_{ij}-
\frac{1}{2}F_{ijk}F_{ijk}+2\Pi_{ij}\partial_i B_{0j}+\partial_jB_{ji}\Pi_{oi}
\right.\right.\nonumber\\
&-&\left.\left.\partial_i{\bar{\cal{C}}}_j\left(\partial_i{\cal{C}}_j-\partial_j{
\cal{C}}_i\right)-{\bar{\cal{P}}}_i{\cal{P}}_i-\partial_i{\cal{P}}_i{\cal{C}}_0+{\cal{P}}_0
\partial_i{\cal{C}}_i+{\bar{\cal{P}}}_i\partial_i
{\bar{\cal{C}}}_0-{\bar{\cal{P}}}_0\partial_i{\bar{\cal{C}}}_i
\right.\right.\nonumber\\
&-&\left.\left.\partial_i\bar{\beta}\partial_i \beta +\Pi_{\varphi_1}\partial_i B_{0i}-
\partial_i\varphi_1\Pi_
{0i}-\frac{1}{2}\Pi_{0i}\Pi_{0i}-\frac{1}{2}\Pi_{\varphi_1}^2 
-{\cal{P}}_0{\bar{\cal{P}}}_0 +\Pi_\beta \Pi_{\bar{\beta}}\right\}\right],\label{zpsi}
\end{eqnarray}
which is invariant under the BRST transformation given in Eq.(\ref{qbrst}) and $D\chi $ is 
the Liouville measure over the entire phase space.
Integrating Eq.(\ref{zpsi}) over ${\cal{P}}_i,{\bar{\cal{P}}}_i,\Pi_\beta$ and $ 
\Pi_{\bar{\beta}}$ 
, we obtain
\begin{eqnarray}
{\cal{P}}_i&=&-\left( \partial_0{\bar{\cal{C}}}_i-\partial_i{\bar{\cal{C}}}_0
\right),\quad\quad \Pi_\beta=\dot{\bar{\beta}},\nonumber\\
{\bar{\cal{P}}}_i&=&\left(\partial_0{\cal{C}}_i-\partial_i{\cal{C}}_0\right),
\quad \quad \quad\;\Pi_{\bar{\beta}}=\dot{\beta},
\end{eqnarray}
after which the generating functional in Eq. (\ref{zpsi}) becomes
\begin{eqnarray}
Z_\Psi^\prime &=&\int D\chi^\prime \exp \left[i\int d^4x \left\{ \Pi^{ij}{\dot{B}}_{ij}-\Pi_
{ij}\Pi_{ij}+2
\Pi_{ij}\partial_i B_{0j}-\frac{1}{12}F_{ijk}F_{ijk}-\frac{1}{2}\Pi_{0i}^2-\frac{1}{2}\Pi_{
\varphi_1}^2\right.\right.\nonumber\\
&+&\left.\left.\Pi^{0i}\left(\partial_0 B_{0i}-\partial_j
 B_{ji}+\partial_i\varphi_1 \right )+\Pi_{\varphi_1}{\left(\partial_iB_{0i}+
\partial_0\varphi_1\right)}-\partial_i{\bar{\cal{C}}}_j\left(\partial_i{\cal{C}}_j-
\partial_j{\cal{C}}_i\right) +\partial_0\bar{\beta
}\partial_0\beta\right.\right.\nonumber\\
&-&\left.\left. \partial_i\bar{\beta}\partial_i\beta+\partial_0{\bar{\cal{C}}}_i
\left(\partial_0{\cal{C}}_i-
\partial_i{\cal{C}}_0\right)-\partial_i{\bar{\cal{C}}}_0\left(\partial_0{\cal{C}}_i-\partial_i
{\cal{C}}_0\right)+\partial\cdot{\cal{C}}{\cal{P}}_0+\partial\cdot{\bar{\cal{C}}}{\bar{\cal
{P}}}_0+{\bar{\cal{P}}}_0{\cal{P}}_0 \right\}\right].\nonumber\\
\label{seffp}
\end{eqnarray}
Finally, we integrate over $\Pi_{0i}$, $\Pi_{ij}$ and $\Pi_{\varphi_1}$ to obtain
\begin{eqnarray}
Z_\Psi^{\prime\prime}&=&\int \;D\chi^{\prime\prime}\exp\left[i\int d^4x\left\{\frac{1}{12}F_{
\mu\nu\lambda}F^{\mu\nu\lambda}-\frac{1}{2}
{\left(\partial_\nu B^{\nu\mu}+\partial_\mu\varphi \right)}^2 -\partial_\mu {\bar{\cal{C}}}_
\nu\left(\partial^\mu{\cal{C}}^
\nu-\partial^\nu{\cal{C}}^\mu\right)\right.\right.\nonumber\\
&+&\left.\left.\partial_\mu{\bar{\beta}}
\partial^\mu\beta +\partial\cdot{\cal{C}}{\cal{P}}_0+\partial\cdot{\bar{\cal
{C}}}{\bar{\cal{P}}}_0+{\bar{\cal{P}}}_0{\cal{P}}_0\right\}\right],\label{seffpp}
\end{eqnarray}
which contains the known  free Abelian  4D 2-form effective action \cite{dema,malik1}.
After the integration over ${\cal{P}}_i,{\bar{\cal{P}}}_i,\Pi_\beta$ and $\Pi_{\bar{\beta}}$, 
the set of BRST transformations given in Eq. (\ref{qbrst}) becomes
\begin{eqnarray}
&&\delta_b B_{0i}\;=\left(\partial_0{\cal{C}}_i-\partial_i{\cal{C}}_0\right), \quad\quad 
\delta_b{\cal{C}}_i\;\;=\partial_i\beta,\quad\quad \delta_b{\cal{C}}_0\;=\partial_0\beta,\nonumber\\
&&\delta_b B_{ij}\;=\left(\partial_i{\cal{C}}_j-\partial_j{\cal{C}}_i\right),\quad\quad\delta_b{
\bar{\cal{C}}}_i\;\;=\Pi_{0i},\quad\quad \delta_b{\bar{\cal{C}}}_0\;=\Pi_{\varphi_1},\nonumber\\
&&\delta_b\varphi_1 \;\;=-{\bar{\cal{P}}}_0,\quad\quad\quad\quad\quad\;\;\; \delta_b\bar{\beta}\;\;=-{
\cal{P}}_0,\quad\;\;\delta_b\beta\;\;=0,\nonumber\\
&&\delta_b{\cal{P}}_0\;\;=0,\quad\quad \quad\quad\quad\quad \quad\;\delta_b{\bar{\cal{P}}}_0=0,
\quad\quad\;\;\; \delta_b \Pi_{0i}=0,\nonumber\\
&&\delta_b \Pi_{\varphi_1}=0,\quad\quad\quad\quad\quad\;\;\;\;\;\;\;\;\delta_b\Pi_{ij}=0,\label
{qbrst1}
\end{eqnarray}
with $\Pi_{0i}=\left(\partial_0B_{0i}-\partial_j B_{ji}+\partial_i\varphi_1\right)$ and 
$\Pi_{\varphi_1}=\left(\partial_0B_{0i}+\partial_0\varphi_1\right)$ that leave the actions in 
Eq. (\ref{seffp}) and Eq. (\ref{seffpp}) invariant.

\section{Dual-BRST symmetry}
In this section, we discuss the dual-BRST symmetry obtained from canonical 
transformations of the field variables  which leaves the gauge 
fixing part of the action invariant.
The kinetic part of the 2-form Lagrangian density can be linearized by introducing the 
Nakanishi-Lautrup type of auxiliary field, $H_\mu$ \cite{naoj} and a massless scalar field 
$\varphi_2$
 as
\begin{equation}
{\cal{L}}_0 =\frac{1}{2}H^\mu H_\mu - \frac{H^\mu}{2} \left ( \varepsilon_{
\mu\nu\eta\kappa}\partial^\nu B^{\eta\kappa}+\partial_\mu\varphi_2 \right ),
\end{equation}
from which the following canonical momenta can be found out as
\begin{equation}
\Pi_{ij}=\frac{\partial{{\cal{L}}_0}}{\partial(\partial_0 B_{ij})}=-\frac{1}{2}\epsilon_
{ijk}H_k, \quad \quad \Pi_{\varphi_2}=\frac{\partial{{\cal{L}}_0}}{\partial\dot{\varphi_2
}}=-\frac{H_0}{2}.
\end{equation}
As done in section I, the effective action in Eq. (\ref{seffp}) can be re-expressed as
\begin{eqnarray}
S_{eff}^\prime &=& \int d^4x \left[\Pi_{\varphi_2}\dot{\varphi_2} +\Pi_{ij}{\dot{B}}_{ij}-\Pi_
{ij}\Pi_{ij}-2
\Pi_{ij}\partial_i B_{0j}-\Pi_{\varphi_2}\epsilon_{ijk}\partial_iB_{jk}-\frac{1}{2}\epsilon_
{ijk}\Pi_{jk}\partial_i\varphi_2  \right.\nonumber\\
&+&2\Pi_{\varphi_2}^2+\left.\Pi^{0i}\left(\partial_0 B_{0i}+\partial_j
 B_{ji}-\partial_i\varphi \right )
-\frac{1}{2}\Pi_{0i}^2-\frac{1}{2}\Pi_{\varphi_1}^2+\Pi_{\varphi_1}{\left(\partial_iB_{0i}+
\partial_0\varphi\right)}^2 \right.\nonumber\\
&-&\partial_i{\bar{\cal{C}}}_j\left(\partial_i{\cal{C}}_j-
\partial_j{\cal{C}}_i\right)+\left.\partial_0{\bar{\cal{C}}}_i\left(\partial_0{\cal{C}}_i-
\partial_i{\cal{C}}_0\right)
- \partial_i{\bar{\cal{C}}}_0\left(\partial_0{\cal{C}}_i-\partial_i
{\cal{C}}_0\right)+\partial\cdot{\cal{C}}{\cal{P}}_0+\partial\cdot{\bar{\cal{C}}}{\bar{\cal
{P}}}_0 \right.\nonumber\\
&+&{\bar{\cal{P}}}_0{\cal{P}}_0-\partial_i\bar{\beta}\partial_i\beta+\left. \partial_0\bar{
\beta
}\partial_0\beta\right].\label{seffpr}
\end{eqnarray}
\subsection{Canonical Transformations}
 We make canonical transformations as follows
\\Bosonic sector:
\begin{eqnarray}
&&\Pi_{ij}\;\rightarrow \tilde{\Pi}_{ij}\;=\frac{1}{2}\;\epsilon_{ijk}\;\Pi_{ok},\quad \quad 
\;\Pi_{0i}
\rightarrow \tilde{\Pi}_{0i}=\epsilon_{ijk}\;\Pi_{jk},\nonumber\\
&&B_{ij}\;\rightarrow \tilde{B}_{ij}\;=-\epsilon_{ijk}\;B_{0k},\quad\quad \;B_{0i}\rightarrow \tilde
{B}_{0i}=-\frac{1}{2}\;\epsilon_{ijk}\;B_{jk},\nonumber\\
&&\Pi_{\varphi_1}\rightarrow \tilde{\Pi}_{\varphi_1}= 2\Pi_{\varphi_2}, \quad\quad\;\quad\;\;\;\; 
\varphi_1\;\rightarrow \tilde{\varphi}_1\;=\frac{1}{2}\varphi_2,
\end{eqnarray}
ghost sector:
\begin{eqnarray}
&&{\cal{C}}_i\rightarrow {\cal{C}}_i^\prime \;\;\;={\bar{\cal{C}}}_i,\quad\; {\bar{\cal{C}}}_i
\rightarrow {\bar{\cal{C}}}_i^\prime\;={\cal{C}}_i,\quad \;{\cal{C}}_0\;\rightarrow {\cal{C}}_0^
\prime\;\;={\bar{\cal{C}}}_0, \nonumber\\
&&{\bar{\cal{C}}}_0\rightarrow {\bar{\cal{C}}}_0^\prime \;\;={\cal{C}}_0,\quad \;{\cal{P}}_i
\rightarrow {\cal{P}}_i^\prime={\bar{\cal{P}}}_i,\quad {\bar{\cal{P}}}_i\;\rightarrow {\bar{\cal
{P}}}_i^\prime\;={\cal{P}}_i,\nonumber\\
&&{\cal{P}}_0\rightarrow {\cal{P}}_0^\prime\; ={\bar{\cal{P}}}_0,\quad {\bar{\cal{P}}}_0
\rightarrow {\bar{\cal{P}}}_0^\prime = {\cal{P}}_0,\quad \Pi_\beta\rightarrow 
\Pi_\beta^\prime=\Pi_{\bar{\beta}}, \nonumber\\
&&\Pi_{\bar{\beta}}\rightarrow \Pi_{\bar{\beta}}^\prime =\Pi_\beta, \quad \beta\;\rightarrow 
\beta^\prime\;=\bar{\beta},\quad \;\;\;\bar{\beta}\;\rightarrow \bar{\beta}^\prime\;\;=\beta.
\end{eqnarray}
It can be easily seen that $(\tilde{\Pi}_{ij},\tilde{B}_{ij})$ and 
$(\tilde{\Pi}_{0i},\tilde{B}_{0i})$ satisfy the commutation relations
\begin{eqnarray}
\left[\tilde{B}_{ij}({\bf x}),\tilde{\Pi}^{kl}({\bf y})\right ]&=&\frac{i}{2}\;\left( 
\delta_i^k\delta_j^l -\delta_i^l\delta_j^k \right)\delta^3({\bf x}- {\bf y}),\nonumber\\
\left[\tilde{B}_{0i}({\bf x}),\tilde{\Pi}^{0j}({\bf y})\right ]&=&i\;\delta_i^j\delta^3({\bf 
x}- {\bf y}),
\end{eqnarray}
same as Eq. (\ref{com1}) and Eq. (\ref{com2}). Similarly, 
it is trivial to see that other transformations in the ghost sector and bosonic sector are 
also canonical as they satisfy the anti-commutation/commutation relation given by Eq. (\ref{acom1}) and 
Eq. ( \ref{com3}).
After the canonical transformations, the BRST charge given in Eq.( \ref{brsc}) becomes
\begin{equation}
Q_d=i\int d^3x \left[ -\epsilon_{ijk}\partial_i\Pi_{0j}{\bar{\cal{C}}}_k+\epsilon_
{ijk}{\cal{P}}_i\Pi_{jk}+2\Pi_{\varphi_2}{\cal{P}}_0+{\bar{\cal{P}}}_0\Pi_\beta +{\bar{\cal
{P}}}_i\partial_i\bar{\beta}\right],
\end{equation}
which will generate another form of BRST symmetry as follows
\begin{eqnarray}
&&\delta_d B_{0i}\;=-\epsilon_{ijk}\;\partial_j{\bar{\cal{C}}}_k,\quad\quad\quad \delta_d B_{ij}=-
\epsilon_{ijk}\;{\cal{P}}_k,\nonumber\\
&&\delta_d{\cal{C}}_0\;\; =2\Pi_{\varphi_2},\quad\quad \quad\;\quad\quad\delta_d {\bar{\cal{C}}}_0\;\;=
\Pi_\beta,\nonumber\\
&&\delta_d{\cal{C}}_i\;\; =\epsilon_{ijk}\;\Pi_{jk},\quad\quad\quad\quad\;\delta_d{\bar{\cal{C}}}_i\;\;=
\partial_i\bar{\beta},\nonumber\\
&&\delta_d{\bar{\cal{P}}}_i\;\;=-\epsilon_{ijk}\;\partial_j \Pi_{0k},\quad\quad\; \delta_d{\cal
{P}}_i\;\;=0,\nonumber\\
&&\delta_d\beta \;\;\;=-{\bar{\cal{P}}}_0,\quad\quad\quad\quad\quad\;\delta_d\bar{\beta}\;\;
\;\;=0,\nonumber\\
&&\delta_d\Pi_{\beta}\;=0,\quad\quad\quad\quad\quad\quad\;\;\; \delta_d\Pi_{\bar{\beta}}\;=-\partial_i{\bar{\cal{P}}}_i,
\nonumber\\
&&\delta_d\Pi_{\varphi_1} =0,\quad\quad \quad\quad\quad\quad\;\;\delta_d\varphi_1\;\;=0,\nonumber\\
&&\delta_d\Pi_{\varphi_2}=0,\quad\quad\quad\quad\quad\quad\;\;\delta_d\varphi_2\;\;=-2{\cal{P}}_0,\nonumber\\
&&\delta_d{\cal{P}}_0\;\;=0,\quad\quad\quad\quad\quad\quad\;\;\delta_d{\bar{\cal{P}}}_0\;\;=0,\nonumber\\
&&\delta_d\Pi_{0i}\;=0,\quad\quad\quad\quad\quad\quad\;\;\delta_d\Pi_{ij}\;=0.
\label{dbrst}
\end{eqnarray}
The set of BRST transformations given in Eq. (\ref{dbrst}), when integrated over 
${\cal{P}}_i,{\bar{\cal{P}}}_i,\Pi_\beta$ and $\Pi_{\bar{\beta}}$, becomes
\begin{eqnarray}
&&\delta_d B_{0i}\;\;=-\epsilon_{ijk}\;\partial_j{\bar{\cal{C}}}_k,\quad\quad \delta_d B_{ij}=-
\epsilon_{ijk}\;\;\left(\partial_0{\bar{\cal{C}}}_k-\partial_k{\bar{\cal{C}}}_0
\right),\nonumber\\
&&\delta_d{\cal{C}}_0\;\;\;=2\Pi_{\varphi_2},\quad\quad \quad\quad\;\delta_d {\bar{\cal{C}}}_0
\;\;=\Pi_\beta,\nonumber\\
&&\delta_d{\cal{C}}_i\;\;\;=\epsilon_{ijk}\;\Pi_{jk},\quad\quad\quad\;\delta_d{\bar{\cal
{C}}}_i\;\;=
\partial_i\bar{\beta},\nonumber\\
&&\delta_d\beta \;\;\;\;=-{\bar{\cal{P}}}_0,\quad\quad\quad\quad\;\;\delta_d\bar{\beta}\;\;\;=0,
\nonumber\\
&&\delta_d\Pi_{\varphi_1} =0,\quad\quad \quad\quad\quad\quad\delta_d\varphi_1\;\;=0,\nonumber\\
&&\delta_d\Pi_{\varphi_2}=0,\quad\quad\quad\quad\quad\quad\delta_d\varphi_2\;\;=-2{\cal{P}}_0,
\nonumber\\
&&\delta_d{\cal{P}}_0\;\;=0,\quad\quad\quad\quad\quad\quad\delta_d{\bar{\cal{P}}}_0\;\;=0,
\nonumber\\
&&\delta_d\Pi_{0i}\;=0,\quad\quad\quad\quad\quad\quad\delta_d\Pi_{ij}\;=0.\label{dbrst1}
\end{eqnarray}
The above transformations are nilpotent and leaves the action given in Eq. (\ref{seffpr}) 
invariant. The gauge fixing part of the effective action given in Eq. (\ref{seffpr}) is
\begin{equation}
S_{gf}=\int d^4x \left[\Pi^{0i}\left(\partial_0 B_{0i}+\partial_j
 B_{ji}-\partial_i\varphi \right )
-\frac{1}{2}\Pi_{0i}^2-\frac{1}{2}\Pi_{\varphi_1}^2+\Pi_{\varphi_1}{\left(\partial_iB_{0i}+
\partial_0\varphi\right)}^2 \right],\nonumber\\
\end{equation}
and its variation under the set of BRST transformations given in Eq. (\ref{dbrst1}) 
independently vanishes i.e. $\delta_d S_{gf}=0$. The variation of the kinetic part cancels 
with the variation of the ghost part of the action.
Such a from of BRST symmetry is referred to as ``Dual-BRST symmetry"\cite{hama,guma}.
Hence, we observe that the $Q_b$ and $Q_d$, the generator of the BRST and dual-BRST symmetry respectively, are related through canonical transformations and therefore is not an independent symmetry but the generalization of usual BRST symmetry. 
\section{New form of BRST symmetry}
We make general transformation in the sectors of Lagrange multipliers and their corresponding 
momenta such as $(\Pi^{0i},B_{0i})$, $({\cal{P}}_0,{\cal{C}}_0)$,
 $(\Pi_{\varphi_1},\varphi_1)$ and $({\bar{\cal{P}}}_0,{\bar{\cal{C}}}_0)$ as follows
\begin{eqnarray}
&&\Pi^\prime_{0i}\;=-\Pi_{0i}-\left (\partial_0 B_{0i}+\partial_j B_{ij}\right), \quad \quad \;B^
\prime_{0i}= B_{0i}, \nonumber\\
&&\Pi^\prime_{\varphi_1}=-\Pi_{\varphi}-\partial_i B_{0i}, \quad\quad \quad\quad\quad\quad\quad
\varphi^\prime_1\;=\varphi_1, \nonumber\\
&&{\cal{P}}^\prime _0\; = -{\cal{P}}_0-\left (\partial_0{\bar{\cal{C}}}_0-\partial_i{\bar{\cal
{C}}}_i\right),\quad\quad\quad\;\; {\cal{C}}^\prime_0\;\;={\cal{C}}_0,\nonumber\\
&&{\bar{\cal{P}}}^\prime_0 \;= -{\bar{\cal{P}}}_0-\left(\partial_0{\cal{C}}_0-\partial_i{\cal
{C}}_i\right), \quad \quad \quad\;\;\;{\bar{\cal{C}}}^\prime_0\;\;={\bar{\cal{C}}}_0. \label{getr}
\end{eqnarray}
Since, these transformations have Jacobian=-1, therefore it will not affect the path integral 
measure. The effective action given by Eq.(\ref{seffp}) retains its form but the 
transformations in Eq. (\ref{getr}) will give rise to a set of new form of BRST symmetry as 
follows
\begin{eqnarray}
&&\delta_b B_{0i}\;=\left(\partial_0{\cal{C}}_i-\partial_i{\cal{C}}_0\right), \quad 
\quad\quad\quad \quad\quad\delta_b B_{ij}\;\;=\left( \partial_i{\cal
{C}}_j-\partial_j{\cal{C}}_i\right),\nonumber\\
&&\delta_b\beta \;\;\;= 0,\quad \quad\quad\quad\quad\quad \quad\quad\quad\quad\quad\;\delta_b\Pi_{ij}\;\;=0,\\
\nonumber\\ &&\delta_b{\bar{\cal{C}}}_0\;\;=-\Pi_{\varphi_1}-\partial_i B_{0i},
\quad\quad\quad\quad\quad
\quad\delta_b{\bar{\cal{C}}}_i\;\;\;=-\Pi_{0i}-\left(\partial_0 B_{0i}+\partial_j B_{ij}\right),
\nonumber\\
&&\delta_b{\bar{\beta}}\;\;\;=-{\cal{P}}_0-\left(\partial_0{\bar{\cal{C}}}_0-\partial_i{\bar{\cal
{C}}}_i\right),\quad\quad
\;\;\;\delta_b\varphi_1\;\;=-{\bar{\cal{P}}}_0-\left(\partial_0{\cal{C}}_0-\partial_i{\bar{C}}_i
\right),\nonumber\\
&&\delta_b\Pi_{0i}=\delta_b\left(\partial_0B_{0i}+\partial_jB_{ij}
\right),\quad\quad\quad\quad\;\delta_b\Pi_{\varphi_1}=-\delta_b(\partial_iB_{0i}),\nonumber\\
&&\delta_b{\cal{P}}_0\;=-\delta_b\left(\partial_0{\bar{\cal{C}}}_0-\partial_i{\bar{\cal{C}}}_i
\right),\quad\quad\quad\quad\;\delta_b{\bar{\cal{P}}}_0\;\;=-\delta_b\left(\partial_0{\cal{C}}_0-
\partial_i{
\cal{C}}_i\right).\label{qbrst2}
\end{eqnarray}
The non-trivial set of transformations are given in Eq. (\ref{qbrst2}). These set of BRST 
transformations are also nilpotent and leaves the effective action given in Eq. (\ref{seffp}) 
invariant.

\section{New form of dual-BRST symmetry}
If we make the similar general transformations as given in Eq. (\ref{getr}), we obtain an 
another form 
of dual-BRST symmetry. The new set of dual-BRST transformations obtained are as follows
\begin{eqnarray}
&&\delta_d B_{0i}\;=-\epsilon_{ijk}\;\partial_j{\bar{\cal{C}}}_k,\quad\quad\quad\quad\quad\quad\;\; \delta_d 
B_{ij}=\epsilon_{ijk}\;\left(\partial_0{\bar{\cal{C}}}_k-\partial_k{\bar{\cal{C}}}_0
\right),\nonumber\\
&&\delta_d{\cal{C}}_i\;\;=\epsilon_{ijk}\;\Pi_{jk},\quad\quad\quad\quad\quad\quad\quad\quad\delta_d{\bar{\cal{C}}}_i\;\;=
\partial_i\bar{\beta},\nonumber\\
&&\delta_d{\cal{C}}_0\;\;=2\Pi_{\varphi_2},\quad\quad\quad \quad\quad\quad\quad\quad\quad\delta_d{\bar{\cal{C}}}_0\;=\partial_0\bar{\beta},\nonumber\\
&&\delta_d\bar\beta\;\;\; =0,\quad\quad\quad\quad\quad\quad\quad\quad\quad\quad\;\;\delta_d\varphi_1\;=0,\nonumber\\
&&\delta_d\Pi_{\varphi_2}=0,\quad\quad\quad\quad\quad\quad\quad\quad\quad\quad\; \delta_d\Pi_{ij}=0,\\
\nonumber\\
&&\delta_d\beta\;\;\; ={\bar{\cal{P}}}_0+\left(\partial_0{\cal{C}}_0-\partial_i{\cal{C}}_i\right),\quad\quad\quad\delta_d\varphi_2=2\left({\cal{P}}_0+\partial_0{\bar{\cal{C}}}_0-\partial_i{\bar{\cal{C}}}_i\right),\nonumber\\
&&\delta_d\Pi_{\varphi_1}=-\delta_d(\partial_iB_{0i}),\quad\quad\quad\quad\quad\quad\;\delta_d{\cal{P}}_0=-\delta_d(\partial_0{\bar{\cal{C}}}_0-\partial_i{\bar{\cal{C}}}_i),\nonumber\\
&&\delta_d\Pi_{0i}\;=-\delta_d\left(\partial_0B_{0i}+\partial_jB_{ij}\right),\quad\quad\;\delta_d{\bar{\cal{P}}}_0=-\delta_d\left(\partial_0{\bar{\cal{C}}}_0-\partial_i{\bar{\cal{C}}}_i\right).\label{qdbrst2}
\end{eqnarray}
The non-trivial set of transformations are given in Eq. (\ref{qdbrst2}). These transformation are also nilpotent and leave the effective action given in Eq. (\ref{seffpr}) invariant.
\section{Conclusion}
We consider BRST-BFV formulation for Abelian rank-2 tensor gauge field theory to discuss 
various forms of BRST transformations. Dual-BRST was claimed to be a independent symmetry in 
the literature. We show that dual-BRST symmetry is not an independent symmetry but can be 
obtained from usual BRST by using a canonical transformations in the Bosonic and ghost sector 
of the theory.
\begin{equation}
\delta_b\phi=-i\left[\phi,Q_b\right]\stackrel{Canonical \;Transformations}{--------\longrightarrow}\;\delta_d\phi=-i\left[\phi,Q_d\right]
\nonumber
\end{equation}
There exists a mapping between the de-Rham cohomological operators( exterior derivative d, co-exterior derivative $\delta$ and the operator $\bigtriangleup=\left\{d,\delta\right\} $) of the differential geometry and the generator of various forms of BRST symmetry ($Q_b, Q_d$ and  $Q_w )$ respectively, where the $Q_w$ is the generator of bosonic symmetry \cite{hama,guma}. Therefore, it is interesting to observe that the two de-Rham cohomological operators (d and $\delta)$ can related through canonical tranformations.
 We also consider the general transformations of Lagrange multipliers in bosonic and ghost sector to further obtain  the different forms of BRST and dual-BRST symmetry.
All the forms of BRST symmetry presented in this paper are local, covariant and nilpotent. New forms of BRST transformations can be used in conjunction with its usual form. These different forms of BRST transformations may be useful in technical point of view, in particular these can simplify the renormalizable program. Further, BRST-BFV technique straightforwadly can be extended to higher rank antisymmetric tensor theories. In such cases, the theory will be reducible at higher and higher levels so that more and more ghosts will be required.\\
\\
\noindent
{\Large{\bf {Acknowledgment}}} \\

\noindent
We thankfully acknowledge the financial support
from the Department of Science and Technology (DST), Government of India, under
the SERC project sanction grant No. SR/S2/HEP-29/2007.\\

\end{document}